\documentclass[rnote]{aa} 
\usepackage{graphicx}
\usepackage{rotating}
\usepackage{txfonts}
\usepackage{natbib}
\bibpunct{(}{)}{;}{a}{}{,} 

\pdfoutput=1

\def\subsun{\mbox{$_{\normalsize\odot}$}}
\def\sqdeg{\mbox{${\rm deg}^2$}}
\def\lesssim{\mathrel{\hbox{\rlap{\hbox{\lower4pt\hbox{$\sim$}}}\hbox{$<$}}}}
\def\gtrsim{\mathrel{\hbox{\rlap{\hbox{\lower4pt\hbox{$\sim$}}}\hbox{$>$}}}}

\begin{document}
%
   \title{The ALHAMBRA survey: Discovery of a faint QSO at $z=5.41$}


   \author{I. Matute \inst{1}
          \and
          J. Masegosa \inst{1} 
          \and
          I. M\'arquez \inst{1}
	  \and	
	  A. Fern\'andez-Soto  \inst{2,3}
	  \and
	  C. Husillos \inst{1}
	  \and
	  A. del Olmo \inst{1}
	  \and
	  J. Perea \inst{1}
	  \and
          M. Povi\'c \inst{1}
	  \and
	  B. Ascaso \inst{1}
          \and
	  E.\,J. Alfaro \inst{1}
	  \and
	  M. Moles \inst{1,4}
	  \and
	  J.\,A.\,L. Aguerri  \inst{5}
	  \and  
	  T. Aparicio--Villegas \inst{6}
	  \and
	  N. Ben\'itez \inst{1}
	  \and
	  T. Broadhurst  \inst{7}
	  \and
	  J. Cabrera--Cano  \inst{1,8}
	  \and
	  F.\,J. Castander \inst{9}
	  \and
	  J. Cepa  \inst{5,10}
	  \and
	  M. Cervi\~no  \inst{1,5}
	  \and
	  D. Crist\'obal-Hornillos \inst{1,4}
	  \and
	  L. Infante \inst{10}
	  \and  
	  R.\,M. Gonz\'alez Delgado  \inst{1}
	  \and
	  V. J. Mart\'inez \inst{3,12,13}
	  \and
	  A. Molino \inst{1}
	  \and
	  F. Prada \inst{1}
	  \and
	  J. M. Quintana  \inst{1}
          }

   \institute{Instituto de Astrof\'isica de Andaluc\'ia (CSIC),
              Glorieta de la Astronom\'ia s/n, E--18008 Granada, Spain\\
              \email{matute,\,isabel,\,pepa,\,cesar,\,chony,\,jaime,\,mpovic,\,emilio,\,benitez,\,mcs,\,rosa,\,amb,\,fprada,\,quintana@iaa.es}
	 \and
	    Instituto de F\'isica de Cantabria (CSIC-UC), E--39005, Santander, Spain; fsoto@ifca.unican.es 
	  \and
	    Unidad Asociada Observatori Astron\`omic (IFCA - Universitat de Val\`encia), Valencia, Spain
         \and
	    Centro de Estudios de F\'isica del Cosmos de Arag\'on (CEFCA), E--44001 Teruel, Spain; moles, dch@cefca.es
	 \and	
	    Instituto de Astrof\'isica de Canarias, La Laguna, Tenerife, Spain; jalfonso@iac.es
	 \and
	  Observatório Nacional-MCT, CEP 20921-400, Rio de Janeiro-RJ, Brazil;  villegas@on.br
	 \and
	    School of Physics and Astronomy, Tel Aviv University, Israel; tjb@wise.tau.ac.il
	 \and
	    Facultad de F\'isica. Departamento de F\'isica At\'omica, Molecular y Nuclear. Universidad de Sevilla, Sevilla, Spain; jcc-famn@us.es
	 \and
	    Institut de Ci\`encies de l'Espai, IEEC-CSIC, Barcelona, Spain; fjc@ieec.fcr.es
	 \and	
	    Departamento de Astrof\'isica, Facultad de F\'isica, Universidad de la Laguna, Spain; jcn@iac.es
	 \and
	    Departamento de Astronom\'ia, Pontificia Universidad  Cat\'olica, Santiago, Chile; linfante@astro.puc.cl
	  \and
	    Departament d'Astronom\'ia i Astrof\'isica, Universitat de Val\`encia, Valencia, Spain; vicent.martinez@uv.es
	  \and 
	    Observatori Astron\`omic de la Universitat de Val\`encia, Valencia, Spain
             }

   \date{Received May ??, 2013; accepted July 12, 2013}

 
  \abstract
   {}
  {We aim to illustrate  the potentiality of the Advanced Large, Homogeneous Area, 
  Medium-Band Redshift Astronomical (ALHAMBRA) survey to investigate 
  the high redshift universe through the detection of quasi stellar objects (QSOs) at redshifts larger than 5.}
  {We searched for QSOs candidates at high redshift by fitting an
    extensive library of spectral energy distributions ---including active and non-active galaxy
    templates as well as stars--- to the  photometric database of the ALHAMBRA survey 
    (composed of 20 optical medium-band plus the 3 broad-band $JHK_s$ near-IR filters).}
   {Our selection over $\approx$1 square
   degree of ALHAMBRA data ($\sim$1/4 of the total area covered by the survey), 
   combined with GTC/OSIRIS spectroscopy, has
   yielded the identification of an optically faint QSO at very high
   redshift ($z$ = 5.41). The QSO has an absolute magnitude of $\sim$-24 
   at the 1450\AA\, continuum, a bolometric luminosity of
   $\approx$2$\times 10^{46}$\,erg\,s$^{-1}$ and an estimated black hole mass of
   $\approx$10$^{8}$M$_{\odot}$. This QSO adds itself to a reduced number of
     known UV faint sources at these redshifts. The preliminary derived space
     density is compatible with the most recent determinations of the
     high-$z$ QSO luminosity functions (QLF). This new detection shows how ALHAMBRA, 
     as well as forthcoming well designed photometric surveys, can provide a wealth of information 
     on the origin and early evolution of this kind of objects.}
   {}

   \keywords{cosmology: Observations -- galaxies: active -- galaxies: distances and redshifts
                galaxies: evolution -- galaxies: high-redshift -- quasars: general
               }

   \maketitle
%

\begin{figure*}[!ht]

  \includegraphics[width=6.5cm, angle=90, trim=2cm 2cm 1cm 3cm, clip]{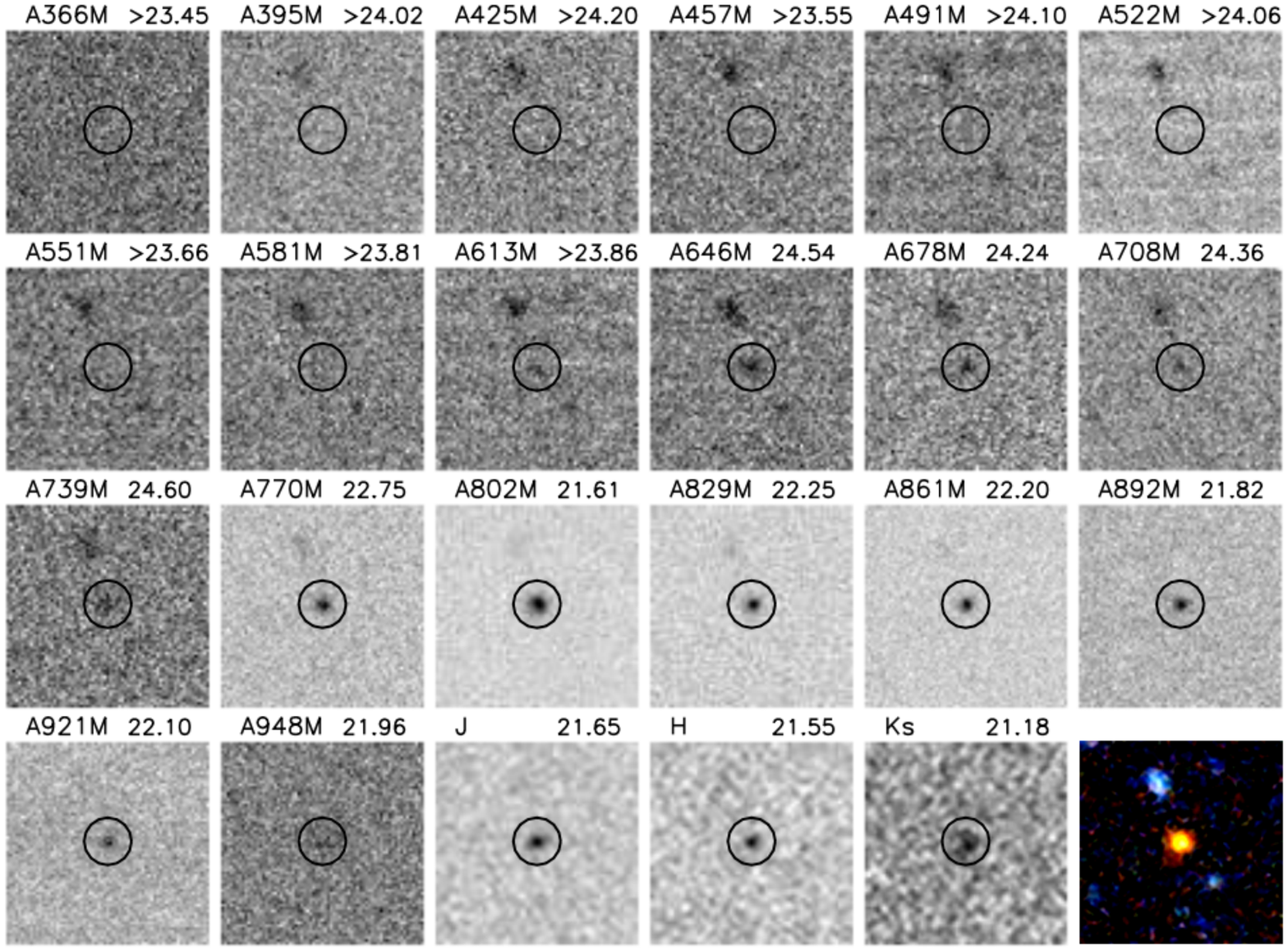}
  \includegraphics[width=9.0cm, trim=0 -1cm 0 0cm, clip]{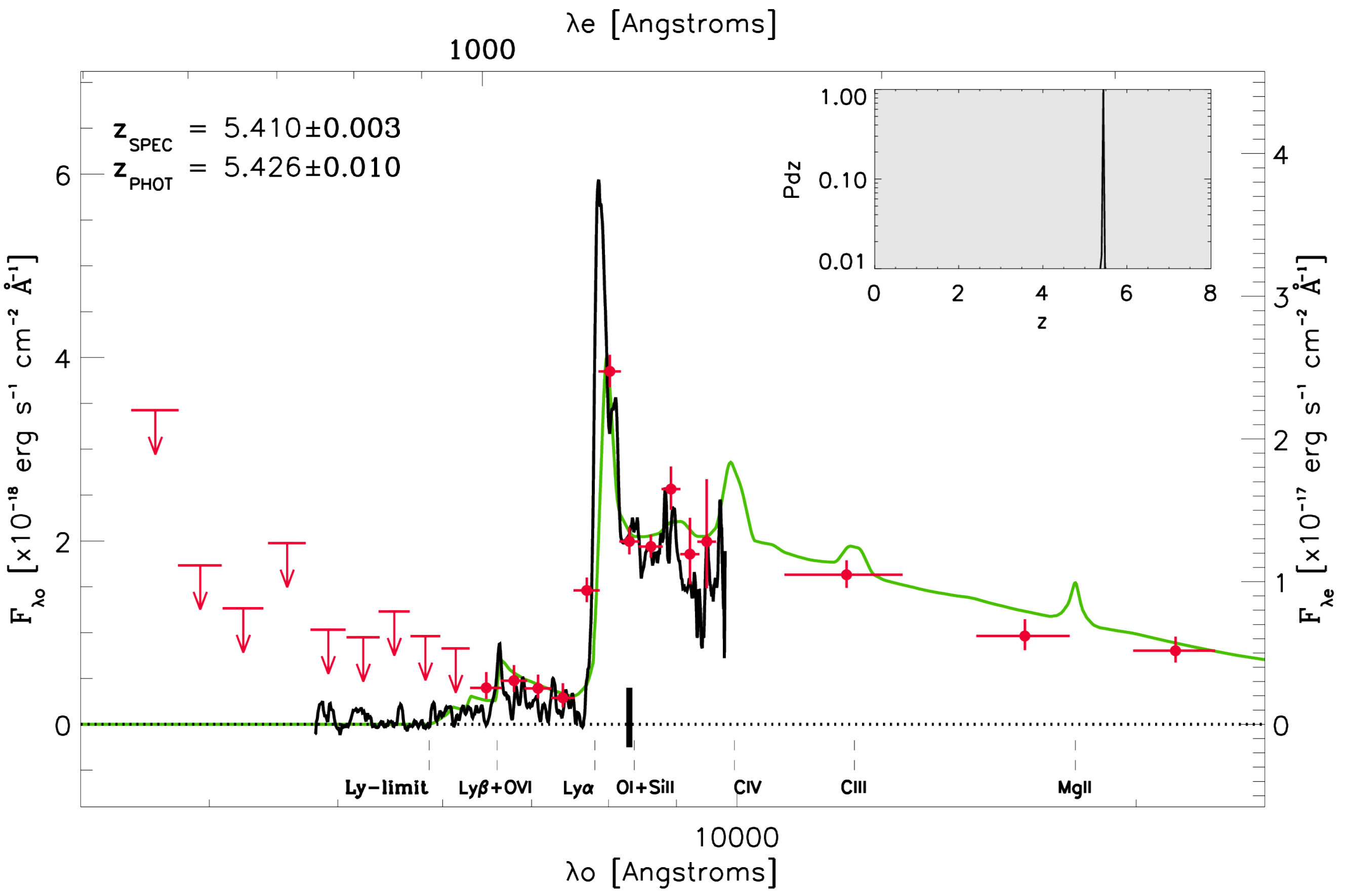}

  \caption{\textit{Left)} Cutouts (15''$\times$15'') in all the
    ALHAMBRA optical/NIR filters for the discovered QSO (highlighted
    by an open circle). Above each cutout we indicate the filter name
    and measured magnitude (or 5-$\sigma$ lower limit). The final
    image is a colour composite of all bands, where the contrast has
    been increased in order to make all objects clearly
    visible. Images are oriented with north up and east to the left.
    \textit{Right)} Optical--NIR spectral energy distribution of the
    discovered QSO.  ALHAMBRA photometric detections are represented as
    circles with associated error bars (arrows indicate the 5$\sigma$
    upper limits). The reference magnitude in the m$_{830}$ filter is
    indicated by a vertical thick line. The best photo-$z$ template
    solution (QSO with optical slope index $\alpha=-0.25$ at
    $z_{\mathrm{PHOT}}=5.426\pm0.010$) is shown as a green line while
    the OSIRIS/GTC spectrum (smoothed with a 7 pixel box) is shown as a
    thick black line, with the redshift probability function (Pd$z$) in the
    inset. The agreement found between the spectro-$z$ and the
    photo-$z$ is remarkable. The most important emission lines for
    QSOs at the redshift of the source are also indicated.}
  \label{fig:High-z-QSO_spectra}
\end{figure*}

\section{Introduction}

It is now widely accepted that the release of gravitational
energy, as matter falls into a supermassive black hole
($\sim$10$^{6-9}$M\subsun; SMBH), is the main generation mechanism for
the high luminosities observed in AGN.  
Inside the variety of the AGN family, quasi-stellar objects (quasars or QSOs)
are the fraction that shows particularly high intrinsic luminosities that
allow them to be detected over very large distances. Consequently, QSOs provide a
way to peer into the physical conditions of an early universe and
study the history of the QSOs and their host galaxy interaction over
cosmic time. The largest samples of known QSOs are provided by the
optical spectroscopic surveys carried out over thousands of
sq.\,degrees of sky (e.g. SDSS and the 2dF QSO survey, 2QZ). These
surveys secure tens of thousand of QSO detections and are able to derive
detailed luminosity functions up to $z\sim3-4$ 
\citep[e.g.,][]{2006AJ....131.2766R,Siana2008,croom09,Ross2012arXiv}
in the form of double power-laws with a characteristic 
luminosity ($M^*$) and faint and bright slopes 
($\alpha$ and $\beta$ for $M>M^*$  and $M<M^*$ respectively).  
The situation at very high-$z$ ($z\gtrsim5$) is quite different. 
At these redshifts, most of the identified QSOs ($\sim$300 objects above 
$z\sim5$ and $\sim$40 at $z>6$) sample only the brightest end of the QSO luminosity function 
\citep[QLF;][]{wolf2003,fan2006,2009AJ....138..305J,2010AJ....139..906W,2012arXiv1212.4493M,Paris2012_SDSS_QSODR9}. 
Over the redshift range $z=[5.0, 6.2]$, less than 10\% of the known QSOs 
\citep[23 objects;][]{Mahabal2005,2009AJ....138..305J,2010AJ....139..906W,Masters2012,2012arXiv1212.4493M}
sample the faint part of the QLF ($M_{1450}>-25$ where $M_{1450}$ is the 
monocromatic luminosity at rest frame 1450\,\AA). 
This not only restrains our knowledge of the true number of
faint QSOs at these redshifts but also severely hampers the derived accuracy of 
the overall QLF (given the high correlation between the LF parameters). 
This uncertainty in the high-$z$ QLF limits our capability to answer some 
important cosmological questions such as the contribution of QSOs to the epoch of 
re-ionization \citep{fan2006,2009ApJS..180..306D,2013MNRAS.428.3058S}, 
the formation of SMBHs within the first billion years of the universe and its 
challenge to models of galaxy formation, BH formation, and BH growth 
\citep{2005Natur.433..604D,2010ApJ...724..915H,Melia2013,Khandai2012}

The optical selection of QSOs has been performed mainly with follow-up
spectroscopic observations of colour--colour selected candidates
\citep[e.g.,][]{croom2004,richards2002,Paris2012_SDSS_QSODR9,2012arXiv1212.4493M,Palanque-Delabrouille2013}, 
using slitless or prism spectroscopic surveys and through poorly efficient flux--limited
spectroscopic surveys \citep[e.g. VIMOS--VLT Deep Survey,][]{bongiorno2007}. 
The novel photometric survey definition by COMBO--17  \citep{wolf2003} 
introduced for the first time a highly efficient selection criteria for QSOs
with photometric redshift (photo-$z$) estimates with precisions of 
$\Delta z$/(1+$z$)$\sim$0.03. Following the
same philosophy, the Advanced Large, Homogeneous Area, Medium-Band
Redshift Astronomical (ALHAMBRA\footnote{http://www.alhambrasurvey.com}) survey
aims at probing a large cosmological fraction of the universe with a
deep and wide catalogue of extragalactic sources with highly accurate
photometry \citep{moles08}.  The survey actually covers
$\sim$4\,\sqdeg over eight different regions of sky and provides a photometric dataset over 20
contiguous, equal-width, non-overlapping, medium-band optical filters
(3500 -- 9700\,\AA) plus the 3 standard broad-band NIR filters
\textit{JHK$_S$} \citep{2009ApJ...692L...5B,2010AJ....139.1242A}.  
The capabilities of the ALHAMBRA survey to select and classify
extragalactic sources with its \textit{very-low-resolution-spectra}
have been explored for galaxies and AGN by Molino et al. (2013, submitted) 
and \citet{matute2012a} respectively. The photometric redshift
precision achieved is $\Delta z$/(1+$z$)$\sim$0.011 and 0.009
for galaxies and AGNs, respectively. A morphological classification for more than
20000 ALHAMBRA galaxies (with photo-$z<1.3$) has been recently derived by 
Povi\'c et al. (2013, submitted).

We present here the discovery of a UV-optically faint QSO at a redshift of 5.41
from the ALHAMBRA photometric survey. The selection criteria and spectroscopic 
observations are described in \S2 and \S3 respectively, while \S4 presents 
the general properties of the QSO. The significance of our discovery is discussed
in \S4 and the conclusions are summarised in \S5. 
Throughout this paper we assume a $\Lambda$CDM cosmology with 
$H_o=73\,\mathrm{km\,s^{-1}}\,\mathrm{Mpc}^{-3}$,
$\Omega_{\Lambda}=0.73$, and $\Omega_{\mathrm{M}}=0.27$. Unless otherwise
specified, all magnitudes are given in the AB system.


\section{High-$z$ QSOs candidates}

Optical $i$-dropouts have proven useful in detecting high-$z$ QSO 
\citep[e.g.,][based on SDSS photometry]{2001AJ....122.2833F}.
Unfortunately, the $i$-dropouts and broad-band colours of a
z$\sim$5--6 QSO can look very similar to those of evolved/passive galaxies at
z$\sim$1 or of local red giants and super-giants.  \citet{matute2012a}
have shown that using the 23 bands of the ALHAMBRA photometric catalogue
it is possible to efficiently classify an important fraction ($\sim$90\%) of the
$z=[0,3]$ BLAGN/QSO population and measure highly accurate photometric redshifts.

We have applied the methodology described by \citet{matute2012a} to
the search for high-$z$ ($z>5$) QSOs using the ALHAMBRA photometric
database. The candidates were selected over $\approx$1 deg$^2$ in the two ALHAMBRA fields 
(out of eight; ALH-2/DEEP2 and ALH-8/SDSS) observable from the 
Roque de los Muchachos observatory in La Palma during the 2011B semester according to
the following criteria:
$i$) located in the fully exposed areas of the images and in regions not contaminated 
by bright sources flux or artifacts (e.g., spikes),
$ii$) detected (at least) in the three NIR bands and the four reddest optical\footnote{Notice that this implies the
detection of flux at $8450< \lambda < 9630$\,\AA, which induces for this sample an upper 
limit on the redshift $z \lesssim 6$ and a photometric upper limit $m_z\lesssim22.5$.} 
filters and $iii$) having a best fit template of a BLAGN/QSO \citep[templates numbers 29 to 59 in][]{matute2012a}, 
and photo-$z\ge5$ with a probability\footnote{The photometric redshift code \textit{LePhare} provides for each source
the normalized probability distribution over redshift (Pd$z$), i.e. at each $z$ (based on the photometry quality, the SED library 
and parameters assumptions) Pd$z$ gives the probability that it is the correct one.} at the selected redshift  $>50$\%. 
As a result of the spectroscopic follow--up of one of the highest probability
candidates, we report the discovery and basic properties of ALH023002+004647, a 
low luminosity and very high redshift QSO. The \textit{left} panel of Fig.\ref{fig:High-z-QSO_spectra} 
shows the ALH023002+004647 cutouts of the 23 ALHAMBRA filters plus a colour composite.

\begin{table}[!t]
\caption{Properties of the quasar ALH023002+004647.}
\label{tab:QSO-params}      
\centering                          
\begin{tabular}{r | l c}       
 \hline\hline 
 ALH-Field     & ALH-2 (DEEP2) & $(i)$\\
 R.A.\,(J2000) & \,\,\,\,02$^{\mathrm{hh}}$\,30$^{\mathrm{mm}}$\,02.27$^{\mathrm{ss}}$  & $(ii)$\\
 DEC.\,(J2000) & +00$^{\mathrm{dd}}$\,46\arcmin\,\,\,\,46.8\arcsec  & $(iii)$\\
 \hline
 $m_{830}$     & $22.25\pm0.08$  & $(iv)$\\
 $z_{\mathrm{PHOT}}$    &  $5.426\pm0.014$  & $(v)$\\
 Best Fit SED  &  qso\_0.25 (model \#42) & $(vi)$\\
 Best Fit E($B-V$)      & 0.08  & $(vii)$\\
 Best Fit Ext. Law      & SMC  & $(viii)$\\
 
 \hline
 $z_{\mathrm{SPEC}}$       & $5.410\pm0.003$  & $(ix)$\\
 EW(Ly${\alpha}$) & 86 & $(x)$\\
 $f_{1450}$       & 1.388\,$\times$\,10$^{-17}$   & $(xi)$\\   
 $M_{1450}$       & $-24.07\pm0.20$   & $(xii)$\\
 $L_{bol}$        & 46.28$^{+0.19}_{-0.35}$    & $(xiii)$ \\
 $M_{BH}$         & (1.52$\pm$0.83)$\times$10$^{8}$ & $(xiv)$ \\
 $\alpha_O$       & 0.25     & $(xv)$ \\
 
\end{tabular}
\tablefoot{\scriptsize 
  ($i$) ALHAMBRA field name; ($ii$,$iii$) Object coordinates; ($iv$) Observed magnitude in the ALHAMBRA A830M filter; 
  ($v$, $vi$, $vii$, $viii$) Derived photo-$z$, template, intrinsic extinction and extinction law from the best fit to the ALHAMBRA photometry
  \citep[for details see][]{matute2012a}; ($ix$) Spectroscopic redshift; 
  ($x$) Rest-frame Ly$\alpha$ equivalent width (\AA) from our modelisation of the spectrum continuum and emission lines (see \S 4.5); 
  ($xi$) Rest-frame monochromatic flux at 1450\,\AA\, in units of erg\,s$^{-1}$\,cm$^{-2}$\,\AA$^{-1}$; 
  ($xii$) Absolute magnitude at 1450\,\AA; 
  ($xiii$) $log_{10}$ of the bolometric luminosity [erg\,s$^{-1}$]; ($xiv$) Black hole mass (M$_{\sun}$) 
  assuming it radiates at the Eddington limit; ($xv$) Optical continuum power law index ($f \propto \nu^{-\alpha}$).}
\end{table}

\section{Observations and Data Reduction}

A pilot program, containing ALH023002+004647, was approved and observed
with the OSIRIS\footnote{http://www.gtc.iac.es/en/pages/instrumentation/osiris.php}
spectrograph at the 10-meter GTC telescope located at the Observatorio
del Roque de los Muchachos in La Palma. Since only a redshift
confirmation and rough spectral classification was required, we used
the low-resolution red grism R300R (with a resolution of 327 and a
wavelength coverage from 5000\,\AA\, to 10000\,\AA) with a slit width of
1.2\arcsec and a 2x2 on-chip binning.  The observations of our
candidates were carried out during September and November 2011 (period
2011B) under spectroscopic conditions.

The reduction process made use of standard IRAF/PyRAF facilities. 
Wavelength calibration was carried out by comparison with exposures of 
HgAr, Xe and Ne lamps.  We used the 5577\,\AA\, [O\,{\tiny I}] sky 
emission line to correct (applying a rigid shift)
for the small offsets that can be introduced by instrument
flexures. Relative flux calibration was carried out by observations of
the spectrophotometric standard stars G158-100 and GRW708247. Each of
the four 1200\,s exposures was reduced and flux calibrated independently
and they were combined afterward. The final spectrum is the result of
4800\,s on--source integration time at a resolution of
6.5\,\AA/pixel with a continuum S/N of 10 at 1450\,\AA\ rest-frame.  The 
spectrum after smoothing with a 7 pixel-wide boxcar filter is shown in
the \textit{right} panel of Fig.\,\ref{fig:High-z-QSO_spectra}, together 
with the ALHAMBRA photometry and best-fit template model.

\begin{figure}
  \centering
  \includegraphics[width=9.0cm]{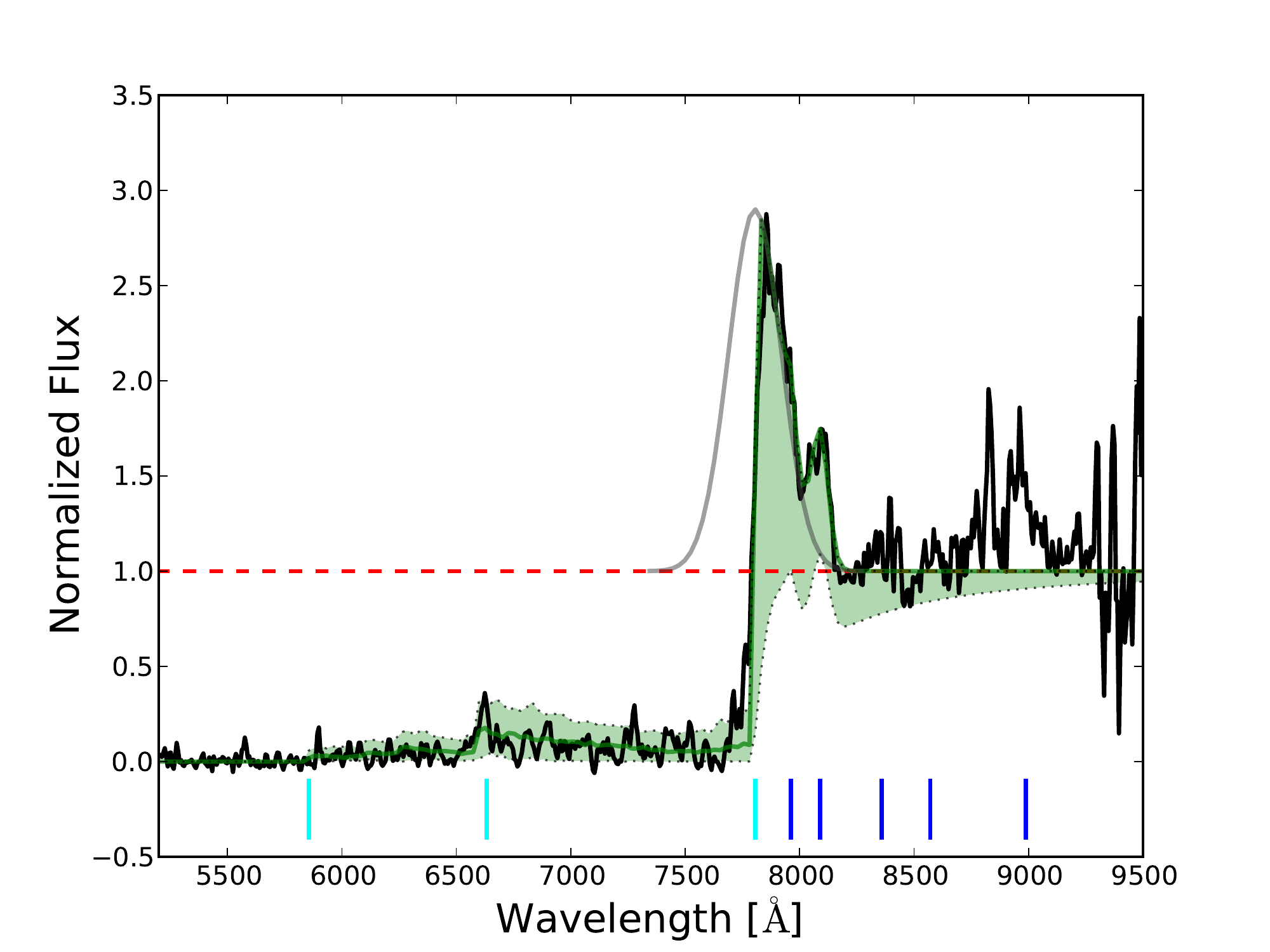}

  \caption{Modelisation of the observed spectrum divided by the fitted
    continuum (\textit{black} line). The Ly$\alpha$, N\,{\tiny V}, and Si\,{\tiny II} $\lambda1260$ lines
    have been modeled with Gaussian profiles. The Gaussian fit to Ly$\alpha$ is shown in grey. 
    The result of dividing the normalized continuum plus the
    modeled lines by the median absorption, as well as upper and lower  2-$\sigma$ deviations, 
    generated by simulating 100 IGM sightlines at the same redshift are shown as a \textit{continuous} 
    (median) and \textit{dotted green} line (2-$\sigma$ limits). 
    Vertical cyan lines mark the positions of the Lyman limit,
    Ly$\beta$, and Ly$\alpha$ lines at the reference redshift. Blue vertical
    lines mark the expected positions of the N\,{\tiny V} $\lambda 1240$, Si\,{\tiny II}
    $\lambda1260$, O\,{\tiny I}$\,\lambda1302$, C\,{\tiny II}\,$\lambda1335$, and
    (O\,{\tiny IV}+Si\,{\tiny IV})$\lambda1400$ emission lines at the same redshift.}
\label{fig:High-z-QSO-linefit}
\end{figure}

\section{Properties of ALH023002+004647}

The spectrum of ALH023002+004647 shows typical features of a high-$z$
QSO spectrum with a a very blue continuum and prominent emission from
broad Ly$\alpha$, N\,{\tiny V}, Ly$\beta$+O\,{\tiny VI} and the
Si\,{\tiny IV}+O\,{\tiny IV]} complex. The continuum blue-wards of
Ly$\alpha$ is substantially reduced due to intervening clouds of
neutral hydrogen present in the IGM at different redshifts along the
line of sight (the Ly$\alpha$ forest). Nevertheless, the presence of a
detectable flux blueward of Ly$\alpha$ is a clear sign that the
universe was highly ionised at the rest-frame of the source. 

Table \ref{tab:QSO-params} resumes the main characteristics of the 
identified QSO.

\subsection{Redshift}

We derived a preliminary spectroscopic redshift $z=5.410\pm0.003$ 
based on the position of the O\,{\tiny I} + Si\,{\tiny II} emission 
line $\lambda$1302. The observed wavelength of Ly$\beta$+O\,{\tiny VI} 
is compatible with this redshift. 

An alternative approach for measuring the redshift, which allows to 
use the highly asymmetric Ly$\alpha$ + N\,{\tiny V} complex,  consists in 
modelling the continuum and the strongest emission lines (Ly$\alpha$, 
N\,{\tiny V} and Si\,{\tiny II}), together with a stochastic realisation of the IGM 
absorption. The result of such an approach is shown in Figure\,\ref{fig:High-z-QSO-linefit}. 
The continuum has been normalised as a power law of slope 2.5 (as 
F$_\lambda \propto \lambda^{-2.5}$). The emission lines have been
modeled with single Gaussian profiles, at a common redshift
$z=5.42$ for Ly$\alpha$ and N{\tiny V}, and $z=5.41$ for Si{\tiny II}
$\lambda$1260. The FWHM of the fitted Ly$\alpha$ is $\approx$1600 
km/s, well above the $\approx$1000 km/s limit offered by the instrumental 
resolution of the grating we used, and usual for this type of objects.

In order to include the effect of the IGM in the spectrum, we have 
generated 100 Ly$\alpha$ forest absorption spectra for lines of sight 
corresponding to the given redshift $z=5.42$, using the technique 
presented in \citet{2003MNRAS.342.1215F}. These spectra are able to 
reproduce the average absorption in the Ly$\alpha$ and Ly$\beta$ 
regions, as well as the incidence of Lyman limit and Damped 
Ly$\alpha$ systems and the basic statistical properties of the
absorption. For each pixel in the 100 resulting spectra we calculated 
the resulting median, $1\sigma$ and $2\sigma$ statistical absorption, 
and used them to generate the IGM absorption component over our 
continuum plus emission-line model.

As can be seen in Figure\,\ref{fig:High-z-QSO-linefit}, the model 
perfectly reproduces the highly asymmetric (Ly$\alpha$ + N\,{\tiny V}) profile, 
as well as the average observed flux in the Ly$\alpha$ and Ly$\beta$ 
forest ranges. Only for the 2.5\% densest IGM absorption sightlines 
the model would underestimate the data redward of Lyman $\alpha$, 
and this is due to the putative presence, in those cases, of a 
strong Damped Ly$\alpha$ system at a redshift close to that of the 
quasar itself--a possibility which is clearly discarded by the data.

The Ly$\alpha$ redshift derived with this second method, $z = 5.420 
\pm 0.005$, is in agreement with the O\,{\tiny I} + Si\,{\tiny II} 
estimate within the errors, but we cannot reject the possibility of 
an offset in the emission-line velocities, as observed in other quasar 
spectra. We find a small discrepancy between the redshift provided 
by Ly$\alpha$ and N\,{\tiny V}, and the low-ionisation emission
line. The disagreement is only of the order of $\sim$500 km/s, but 
in a direction opposite to the expected one. We find the low 
ionisation lines---supposedly marking the real, systemic redshift---to 
be blueshifted with respect to our only clearly detected high ionisation 
line, N\,{\tiny V}, which is in a position compatible with Lyman 
$\alpha$.

As a comparison, \citet{1982ApJ...263...79G} in his pioneering work 
found an offset of $\approx 600$ km/s in the opposite direction to 
the one we find, whereas \citet{2002AJ....124....1R} and 
\citet{2007AJ....133.2222S}, using large samples of quasars, found 
deviations of the same order, albeit with large scatter, covering 
from -500 out to 2000 km/s and beyond. Such scatter would, in fact, 
render the difference observed in ALH023002+004647 compatible with 
their extreme cases. However, we would need a better spectrum 
(particularly in terms of signal-to-noise ratio) to study the 
emission lines in more detail.

\subsection{Companion galaxies}

The cutout images of the QSO (Fig.\,\ref{fig:High-z-QSO_spectra}) show
two companions located 5\arcsec\,to the NE and 4\arcsec\,to the SW. 
According to the ALHAMBRA catalogue (Molino et al. 2013, in prep.), the NE
source corresponds to an $AB(I)=23.8$ blue galaxy at photometric
redshift $z_{ph}=2.25$ whose spectrum peaks at $\approx$ 6500\,\AA, and
the SW source corresponds to an $AB(I)=25.2$ blue galaxy at
photometric redshift $z_{ph}=2.10$ peaking at $\approx$ 5500\,\AA. The
former has a counterpart in the NED database, the source
RCS-01020502096 from the Red-Sequence Cluster Survey \citep{2005ApJS..158..161H}
with a reported redshift of 0.249. However, the $R$ band
magnitude of this object ($R=23.6$) puts it well below the limit that
\citet{2005ApJS..158..161H} quote for good quality redshifts in their
catalogue. No other counterpart has been found for the second
source. According to our data both sources correspond to
low-to-average mass galaxies ($\log (M/M_\odot) \approx 10.5\sim10.8$).

Given the estimated masses for the companions and their distance to the QSO, 
they should provide negligible gravitational lensing amplification
to the quasar image \citep{2005ApJ...626..657W}.  We have checked this for
the observed configuration (including both redshift possibilities in
the case of the NE source), establishing an upper limit to the
possible magnification of 0.4\%.
This may not be exact as the colours of both
sources are intrinsically blue and we have used a simple model of an early 
galaxy type halo---however, no realistic halo model
would yield a significantly different result, not over a factor of two
in the magnification effect \citep{2001PhR...340..291B}.

\subsection{Black hole mass}

The best black hole measurements (apart from reverberation mapping)
can only be obtained using the FWHM of H$\beta$ or Mg{\tiny
  II}\,$\lambda$2800 as a surrogate
\citep{2012NewAR..56...49M}. Unfortunately, there is no currently available 
NIR spectra that covers the wavelength range where these lines are located
(1.8 and 3.1\,$\mu$m for Mg{\tiny II} and H$\beta$
respectively). Nevertheless, a crude estimation of the the black hole
mass and bolometric luminosity can be obtained through the rest-frame
1350\,\AA\ continuum flux measurement. The observed
1350\,\AA\ continuum flux value of
(1.78$\pm$0.27)\,$\times$\,10$^{-18}$\,erg\,s$^{-1}$\,cm$^{-2}$\,\AA$^{-1}$
translates into an absolute luminosity of (5.03$\pm$0.76)\,$\times 10^{45}$  
erg\,s$^{-1}$ at 1350\,\AA, assuming an isotropic emission and that the QSO flux has
not been magnified by gravitational lensing.
A bolometric luminosity of
(1.91$\pm$1.05)\,$\times$10$^{46}$\,erg\,s$^{-1}$ was obtained
following the prescription given in \citet{2006ApJS..166..470R} 
(a $3.8\pm2.0$ factor at $log(\nu) \approx 15.3$\,Hz). 
A black hole mass of (1.52$\pm$0.83)$\times$10$^{8}$M$_{\odot}$ is derived
assuming that the QSO is radiating at the Eddington limit \citep{Perterson1997}.  
We note that a recent study of high-redshift quasars by
\citet{2010AJ....140..546W} has shown that assumption to be
reasonable, with all nine $z\approx 6$ quasars in their sample having
$0.3 < L_{\rm Bol}/L_{\rm Edd} < 2.5$ and $\langle L_{\rm Bol}/L_{\rm
  Edd} \rangle =1.3$.

\subsection{Ancillary data}

  There is a  detection of ALH023002+004647 by SDSS in the Stripe 82 coadd
  (SDSS\,J023002.28+004646.8) with reported magnitudes of 29.00$\pm$1.45, 26.62$\pm$1.00, 24.31$\pm$0.18, 
  22.64$\pm$0.07 and 21.79$\pm$010 in the $u$, $g$, $r$, $i$ and $z$ bands respectively.
  No optical, ultraviolet and radio detections by HST, GALEX,
  XMM-OM and VLA-FIRST are reported by MAST\footnote{\textit{http://archive.stsci.edu/}} for
  ALH023002+004647. The Infrared Science Archive
  (IRSA\footnote{\textit{http://irsa.ipac.caltech.edu}}) provides
  counterparts above 5-$\sigma$ from Spitzer/IRAC observations at 3.6,
  4.5 and 5.8 $\mu$m with a 3.8\arcsec aperture flux of 9.56($\pm0.36$), 13.45($\pm0.57$)
  and 14.17($\pm2.60$) $\mu$Jy,  respectively. These fluxes are compatible, within the errors, with
  the photo-$z$ best fit solution SED. The sky position of ALH023002+004647 is covered by the
  Field-4 of the DEEP2 Redshift Survey. Neither the discovered QSO nor
  the companions have spectroscopic information in the latest data
  release DR4 \citep[][]{Newman2012}.

  The cross-correlation of the ALH023002+004647 position with the high energy
  HEASARC\footnote{\textit{http://heasarc.gsfc.nasa.gov/}} and 
  XSA\footnote{\textit{http://xmm.esac.esa.int/xsa/}} databases reveals that the field 
  has only been observed by \textit{XMM-Newton} with the Epic camera. We find no 
  counterpart associated with the QSO as the closest  \textit{XMM-Newton} detection 
  is located 12\arcsec\, from the QSO and coincides with a galaxy at $z=0.73$.

\section{Discussion}

\begin{figure}[t]
  \centering
  \includegraphics[width=9.2cm]{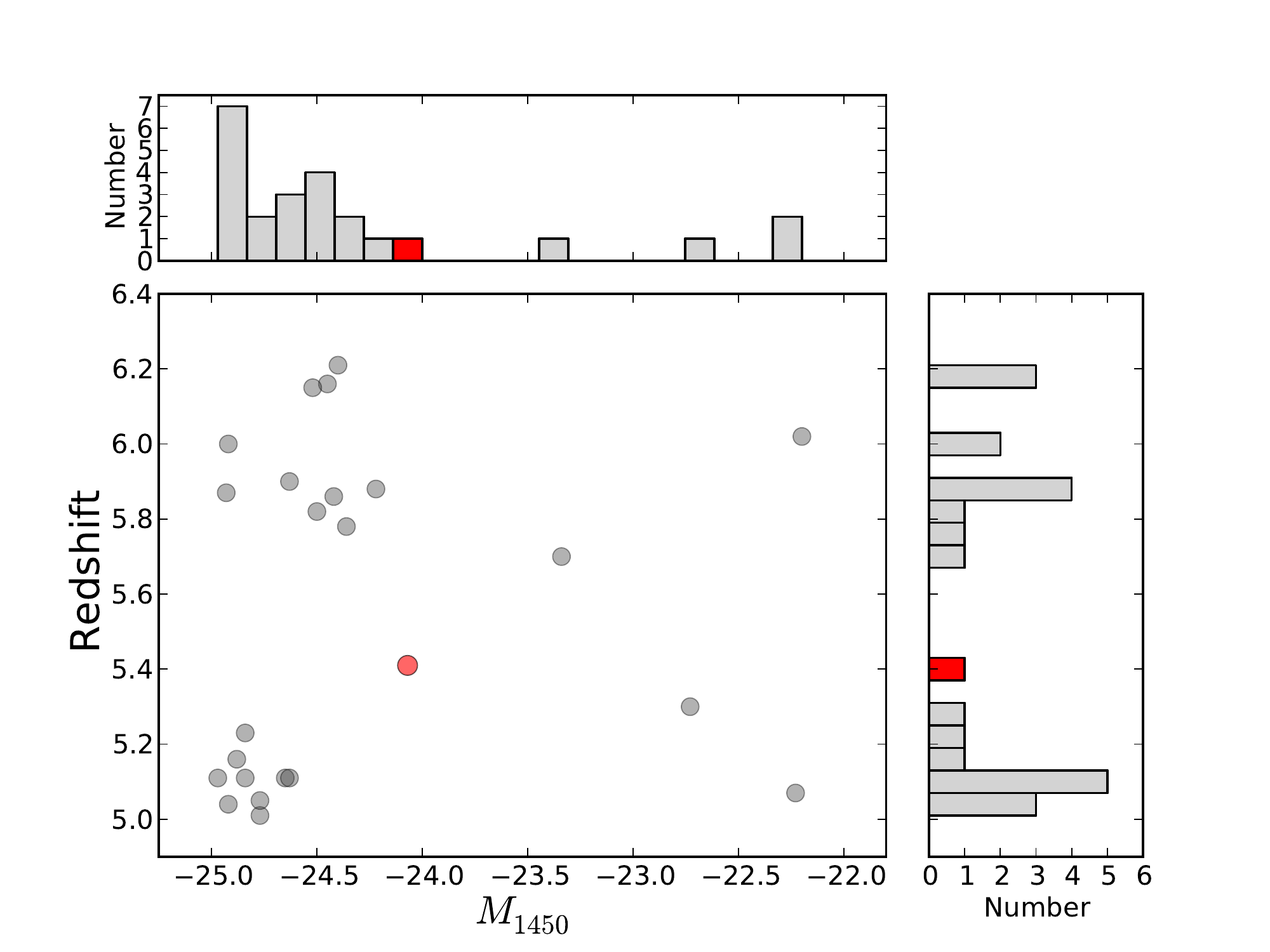}
  \caption{Luminosity--redshift distribution for the 24 currently known faint ($M_{1450}>-25$) QSOs above $z=5$.
  References for the 23 \textit{grey-filled circles} are: \citet[1 src]{Mahabal2005},
  \citet[4 srcs]{2009AJ....138..305J}, \citet[6 srcs]{2010AJ....139..906W}, \citet[2 srcs]{Masters2012} and 
  \citet[][10 srcs]{2012arXiv1212.4493M}. ALH023002+004647 is indicated by a \textit{red-filled circle} and
  \textit{histograms}.}
  \label{fig:ZvsM1450}
\end{figure}

With $M_{1450}=-24.07\pm0.2$, ALH023002+004647 is the 5th (out of 24) 
less luminous QSOs known above redshift of $z=5$. As a consequence, 
it samples the faint part ($M_{1450} > -25$) of QLF, which is 
currently poorly constrained at these redshifts. This is illustrated in 
Fig.\,\ref{fig:ZvsM1450} where we have plotted, in the redshift--luminosity 
space, all the faint spectroscopically confirmed QSO with $z>5$.
  Stronger constraints to
  the space density of high-$z$ low luminosity QSOs are needed to
  address questions as the implications for early AGN/host-galaxy
  interactions or the true fraction of ionizing photons coming from
  AGN activity in an epoch of increasing QSO activity and significant
  BH growth. Therefore, in the following we derive the space density 
  associated  with the ALH023002+004647 detection.

\begin{figure}[t]
  \centering
  \includegraphics[width=9.6cm]{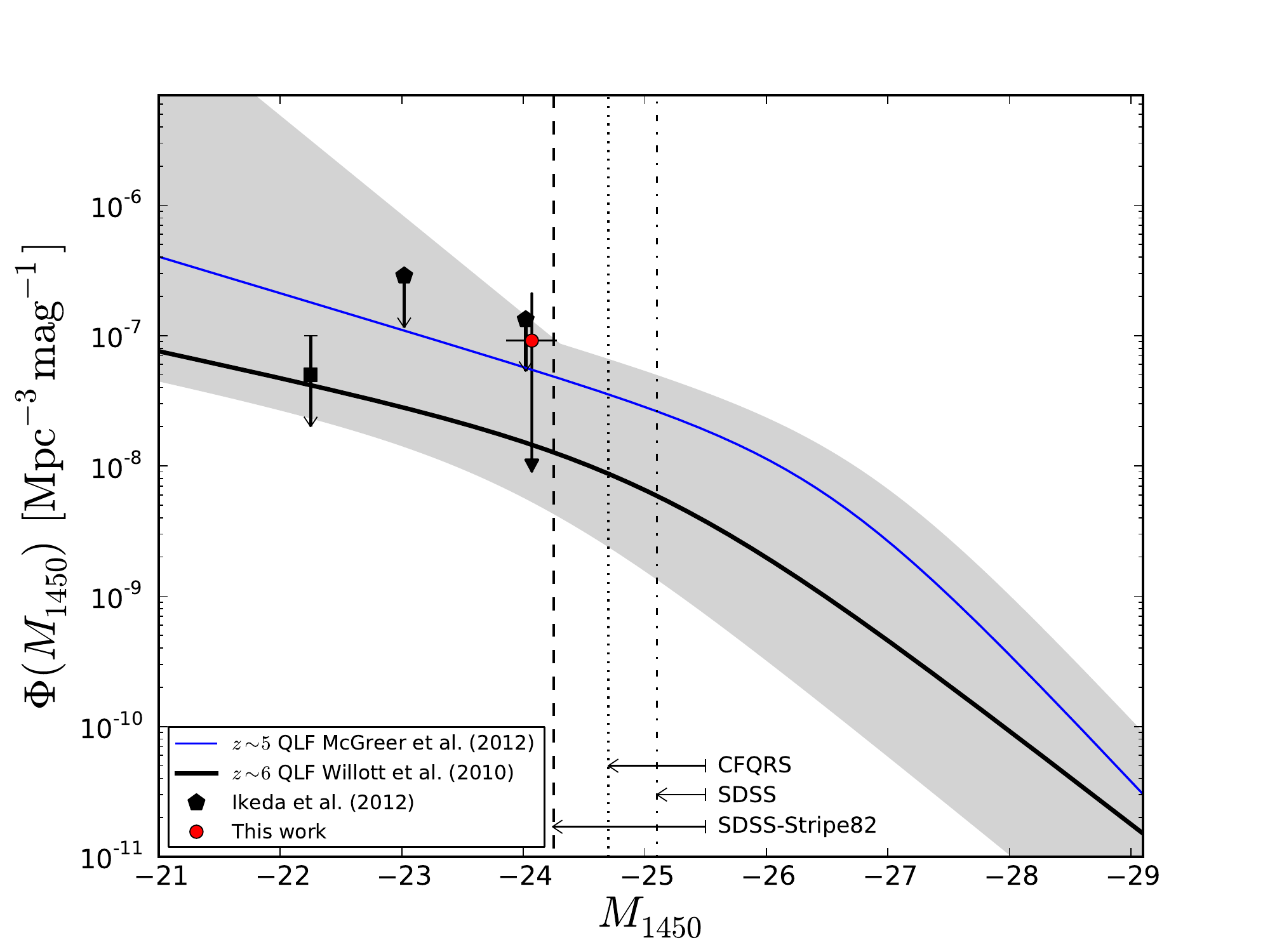}
  \caption{QSO space density distribution as a function of the rest-frame 1450\,\AA\,\, 
  monochromatic magnitude. The corresponding space density of ALH023002+004647 is represented 
  by the red dot. For comparison with some recent determinations of the high-$z$ QLF, we have 
  plotted the $\langle z \rangle \sim 5$  SDSS-Stripe82 QLF derived by 
  \citet[][\textit{blue line}]{2012arXiv1212.4493M} and the $\langle z \rangle \sim 6$ 
  CFQRS+SDSS-Deep QLF derived by \citet[][\textit{black line}]{2010AJ....139..906W}. 
  The vertical lines  indicate the faintest magnitude bin sampled by the SDSS-Main, 
  SDSS-Stripe82 and CFQRS. The grey shaded area delimits the range covered by the 
  published QLFs in the redshift bin [5-6] (see text for details). The \textit{black pentagons}
  represent the upper limits derived by \citet{Ikeda2012} in the COSMOS field while the 
  \textit{black square} shows the space density associated to the faintest known 
  QSO above $z=5$ \citep{2010AJ....139..906W}.}
  \label{fig:High-z-QSO_density}
\end{figure}

  A first approximation to the contribution of ALH023002+004647 to the 
  $z\approx5.5$ space density has been computed using the binned
  1/V$_a$ method of \citet{1980ApJ...235..694A}, where V$_a$ is the comoving
  volume per magnitude and redshift interval accessible to the source,
  taking into account the characteristics of the survey. It is defined
  as,

$$V_a = \int_{\Delta M}\int_{\Delta z} p(M,z)\frac{dV}{dz}\,dz\,dM,$$

  \noindent where $p(M,z)$ is the function that corrects for the
  survey incompleteness at each $M$ and $z$. The space density associated
  to ALH023002+004647 is computed as $\Phi(M,z) = 1/V_a$. The first release
  of the ALHAMBRA photometric database has been recently finalised and its
  completeness needs to be addressed through extensive simulations.
  So, for the time being, and given our first order approximation, we will
  consider that all ``possible'' type-I QSOs within the magnitude limits of
  the survey could have been detected (i.e. $p(M,z)=1$). We find this to be a 
  reasonable assumption based on previous results for QSO completeness maps
  of similar surveys like COMBO-17 \citep[see Fig. 4 in][]{wolf2003}.  Therefore,
  the volume accessible to our QSO is only delimited by the redshift range
  $z$=[5.0, 6.1].  At $z\sim6.1$ the QSO would no longer satisfy our selection
  criteria, as the magnitude $m_{830}$ would be fainter than $\sim25$
  when it samples the Ly-forest of the spectra. The lower limit redshift is fixed by
  our requirement to have a best fit model with $z_{PHOT}\ge5.0$.
  The necessary $K$-corrections have been
  derived using the best-fit template for the photo-$z$ solution.
  Considering the covered area ($\approx$1\,deg$^2$) and the magnitude
  limit of our selection, we derive a space density 
  of $(9.15^{+21.06}_{-9.15})\times10^{-8}\,\mathrm{Mpc}^{-3}\,\mathrm{mag}^{-1}$
  for ALH023002+004647. The errors were estimated from Poisson statistics
  \citep{1986ApJ...303..336G}.  The result is presented in
  Figure\,\ref{fig:High-z-QSO_density}. In order to illustrate the large uncertainties
  of the high-$z$ QLF, we have delimited by a grey-shaded area the luminosity-density 
  space covered by several published QLFs in the $z\sim5-6$ redshift range, namely:
  CFQRS+SDSS-Deep \citep{2010AJ....139..906W}, SDSS-stripe82 \citep{2012arXiv1212.4493M}, 
  COMBO-17 \citep{wolf2003} and SDSS-Deep-Stripe 
  \citep{2009AJ....138..305J}. Two of the most recent determinations of the QLF at $z=5$ 
  \citep{2012arXiv1212.4493M} and $z=6$ \citep{2010AJ....139..906W} are highlighted by 
  the blue and black lines respectively.  We also report the upper limits derived by 
  \citet{Ikeda2012} from the non-detections of QSOs in the redshift bin [4.5--5.5] of the COSMOS
  field (\textit{black pentagons}) and the faintest QSO identification above $z=5$ in the 
  SXDS field (\textit{black square}) as the only constraints to the very faint end of 
  the QLF.
  
The associated error due to  poor statistics hardly constrains the 
$z\sim5-6$ QLFs, making our measurement compatible  with all the previous, 
both single and double power-law, determinations. Nevertheless, the result is 
very promising considering that the QSO selection was made in a fraction of the
total area covered by ALHAMBRA and based on preliminary photometry. With the final 
release of the $\sim$4 sq. degrees of ALHAMBRA data already in place, our collaboration aims 
to further constraint the QLF over a much larger range of magnitudes and redshifts.

\section{Conclusions}
We report the discovery of ALH023002+004647, a new, intrinsically faint, QSO at $z = 5.41$. 
The candidate was selected based on a SED fitting to the 23 bands of the ALHAMBRA 
photometry  and  spectroscopically confirmed with GTC/OSIRIS. The observed 
continuum luminosity at 1450\,\AA \,\,($M_{1450}\sim -24$) and derived bolometric luminosity
of $\approx$2$\times 10^{46}$\,erg\,s$^{-1}$, makes ALH023002+004647 one of the faintest QSOs
discovered above $z=5$.  Based on the ALH023002+004647 space density (and associated errors),  
no tighter constraints can be placed on the $z\sim5-6$ QLF. Nonetheless, we have demonstrated the 
capabilities of the ALHAMBRA survey to select this type of sources. 
An analysis similar to the one presented here, applied to the final release of 
the ALHAMBRA photometric database (DR4) in the near future, will surely provide a significant 
contribution to the QLF at these very high redshifts.

\begin{acknowledgements}
      Part of this work was supported by Junta de Andaluc\'ia, through
      grant TIC-114 and the Excellence Project P08-TIC-3531, and by
      the Spanish Ministry for Science and Innovation through grants
      AYA2006-1456, AYA2010-15169 and AYA2010-22111-C03-02, and
      Generalitat Valenciana project Prometeo 2008/132. IM thanks 
      Cecile Cartozo for the careful reading of the manuscript. This research
      has made use of the NASA/IPAC Extragalactic Database (NED) which
      is operated by the Jet Propulsion Laboratory, California
      Institute of Technology, under contract with the National
      Aeronautics and Space Administration. Based on observations
      collected at the Centro Astron\'omico Hispano Alem\'an (CAHA) at
      Calar Alto, operated jointly by the Max-Planck Institut f\"ur
      Astronomie and the Instituto de Astrof\'{\i}sica de
      Andaluc\'{\i}a (CSIC), and on observations made with the Gran
      Telescopio Canarias (GTC), installed in the Spanish Observatorio
      del Roque de los Muchachos of the Instituto de Astrof\'{\i}sica
      de Canarias, on the island of La Palma. MP acknowledges
      financial support from the JAE-Doc program of the Spanish National 
      Research Council (CSIC), co-funded by the European Social Fund.
\end{acknowledgements}

\bibliographystyle{aa}
\bibliography{mi_bibliografia}
\end{document}